\documentclass[a4paper]{article}

\usepackage{amsfonts}

\def\be{\begin{equation}}
\def\ee{\end{equation}}
\def\bea{\begin{eqnarray}}
\def\eea{\end{eqnarray}}
\def\({\left(}
\def\){\right)}
\def\<{\left<}
\def\>{\right>}

\def\>{\rangle}
\def\<{\langle}
\def\|{\mid}

\def\tr{{\mbox{tr}}}
\def\be{\begin{equation}}
\def\ee{\end{equation}}
\def\bea{\begin{eqnarray*}}
\def\eea{\end{eqnarray*}}
\def\ben{\begin{eqnarray}}
\def\een{\end{eqnarray}}
\def\({\left(}
\def\){\right)}
\def\<{\left<}
\def\>{\right>}

\def\[{\left[}
\def\]{\right]}

\def\+{\bar}
\def\mb{\mathbb}
\def\tr{{\mbox{tr}}}

\def\L{{\cal{L}}}

\def\t{\tilde}

\def\t{\widetilde}

\def\cepsilon{\varepsilon}
\def\cepsilon{{\cal{E}}}

\def\A{{\cal{A}}}
\def\B{{\cal{B}}}

\def\O{{\cal{O}}}

\def\R{{\cal{R}}}

\def\ee{\underline{e}}

\def\aalpha{\underline{\alpha}}
\def\bbeta{\underline{\beta}}
\def\ggamma{\underline{\gamma}}

\begin{document}
\setlength{\unitlength}{1mm}

\pagestyle{empty}
\vskip-10pt
\vskip-10pt
\hfill 
\begin{center}
\vskip 3truecm
{\Large \bf
Five-dimensional SYM from undeformed ABJM}\\
\vskip 2truecm
{\large \bf
Andreas Gustavsson\footnote{a.r.gustavsson@swipnet.se}}\\
\vskip 1truecm
{\it  Center for quantum spacetime (CQUeST), Sogang University, Seoul 121-742, Korea}\\
and\\
\it{School of Physics \& Astronomy, Seoul National University, Seoul 151-747 Korea}
\end{center}
\vskip 2truecm
{\abstract{We expand undeformed ABJM theory around the vacuum solution that was found in arxiv:0909.3101. This solution can be interpreted as a circle-bundle over a two-dimensional plane with a singularity at the origin. By imposing periodic boundary conditions locally far away from the singularity, we obtain a local fuzzy two-torus over which we have a circle fibration. By performing fluctuation analysis we obtain five-dimensional SYM with the precise value on the coupling constant that we would obtain by compactifying multiple M5 branes on the vacuum three-manifold. In the resulting SYM theory we also find a coupling to a background two-form.}}

\vfill 
\vskip4pt
\eject
\pagestyle{plain}

\section{Introduction}
In \cite{Terashima:2009fy} a vacuum solution was found in undeformed ABJM theory, simply by solving the equation of motion in a static field configuration. We will denote the four complex scalar fields in ABJM theory as $Z^A$ for $A=1,2,3,4$. We split $A = (a,\dot{a})$ where $a=1,2$ and $\dot{a}=\dot{1},\dot{2}$. The gauge group is $U(N) \times U(N)$ and $Z^A$ are $N\times N$ bifundamental matrices. However, by utilizing the star-product, we can map these matrices into functions living on a fuzzy two-torus. In the large $N$ limit, we have to leading order that the star-product is just the usual product of functions, and the star-commutator has the leading term which is the Poisson bracket. In this limit the solution that was obtained in \cite{Terashima:2009fy} can be presented as $Z^A = T^A$ where 
\bea
T^a &=& x^a e^{i\psi}\cr
T_{\dot{a}} &=& 0
\eea
Here $x^a$ span a two-dimensional plane, and $\psi$ is a coordinate on a circle fiber over this plane. This solution is a three-manifold $M_3$ with metric 
\bea
ds^2 &=& \delta_{ab} dx^a dx^b + r^2 d\psi^2
\eea
where $r = \sqrt{x^a x^a}$. The scalar curvature is $\R = 1/r$ which is singular at $x^a = 0$.

It is true that $M_3$ does not seem to be translationally invariant, but $M_3$ is translationally invariant in the spacetime of ABJM theory. Since a constant shift of the fermions in ABJM theory is an additional supersymmetry of the Lagrangian, usually refered to as a kinematic supersymmetry \cite{Banks:1996nn}, we deduce that the solution is maximally supersymmetric from the point of view of ABJM theory.

However from the M5 brane point of view, where the M5 brane worldvolume is $\mb{R}^{1,2}\times M_3$, it is unclear to us whether we can have maximal supersymmetry. On $M_3$ we can only find two independent Killing vectors (instead of six as would have been the case had $M_3$ been maximally symmetric)
\bea
V_1 &=& x^2 \partial_1 - x^1 \partial_2 + \arctan \frac{x^2}{x^1} \partial_{\psi}\cr
V_2 &=& \partial_{\psi}
\eea
This means that we can not hope to close untwisted $(2,0)$ supersymmetry variations on the M5 brane into Lie derivatives on $M_3$. Henceforth we will only study the bosonic part of the theory, and leave a possible supersymmetric twisted M5 theory for future studies. 
 
A previous work which dealt with the emergence of the D4 brane from undeformed ABJM theory, is \cite{Terashima:2010ji}. In this work a non-commutativity parameter is introduced by hand, and consequently the Yang-Mills coupling depends on a quantity which is not present in ABJM Lagrangian. In this paper we relate the non-commutativity parameter to parameters which are present in ABJM theory. That is, the rank of the gauge group $N$, and the Chern-Simons level $K$.

\section{Triality of BLG theory}
The R-symmetry of BLG theory is $SO(8)$ which has triality relating $8_v$, $8_c$ and $8_c$ representations. The invariant quantities that carry all these representation indices, are the $SO(8)$ gamma matrices,
\bea
\Gamma_I &=& \(\begin{array}{cc}
0 & \Gamma_{I\alpha\dot{\beta}}\\
\Gamma^{I\dot{\alpha}\beta} & 0
\end{array}\)
\eea
Hermiticity of the gamma matrices implies
\ben
\Gamma_{I\alpha\dot{\beta}}^* &=& \Gamma^{I\dot{\beta}\alpha}\label{ordering}
\een
Triality is a collection of six maps that permutes $8_v$, $8_s$ and $8_c$. We will study those trial maps which relate the ABJM and BLG theories. These act on the indices according to
\bea
I &\rightarrow & \alpha\cr
\alpha & \rightarrow & \dot{\beta}\cr
\dot{\beta} &\rightarrow & I
\eea
and its inverse obtained by reversing the directions of the arrows. Under the above triality map the half-gamma matrices transform according to
\bea
\Gamma^{I\dot{\beta}\alpha} \rightarrow \Gamma^{\alpha I \dot{\beta}} \equiv \Gamma_{I\alpha\dot{\beta}}\cr
\Gamma_{I\alpha\dot{\beta}} \rightarrow \Gamma_{\alpha \dot{\beta} I} \equiv \Gamma^{I\dot{\beta}\alpha}
\eea
To describe how the triality map acts on other quantities, we introduce three bosonic quantities $U^{\alpha}$, $V^{\dot{\alpha}}$ and $W_I$ that we subject to the constraints 
\bea
U_{\alpha} &=& \Gamma_{I\alpha\dot{\beta}} W^I V^{\dot{\beta}}\cr
V^{\dot{\alpha}} &=& \Gamma^{I\dot{\alpha}\beta} U_{\beta} W_I\cr
W_I &=& \Gamma_{I\alpha\dot{\beta}} V^{\dot{\beta}} U^{\alpha}
\eea
and 
\bea
U^{\alpha} U_{\alpha} &=& 1\cr
V^{\dot{\alpha}} V_{\dot{\alpha}} &=& 1\cr
W_I W^I &=& 1
\eea
Then the triality maps read
\bea
X_{\alpha} &=& \Gamma_{I\alpha\dot{\beta}} X^I V^{\dot{\beta}}\cr
\psi^{\dot{\alpha}} &=& \Gamma^{I\dot{\alpha}\beta} \psi_{\beta} W_I\cr
\epsilon_I &=& \Gamma_{I\alpha\dot{\alpha}} \epsilon^{\dot{\alpha}} U^{\alpha}
\eea
and the inverse
\bea
X^I &=& \Gamma^{I\dot{\alpha}\beta}X_{\beta}V_{\dot{\alpha}}\cr
\psi_{\alpha} &=& \Gamma_{I\alpha\dot{\beta}} \psi^{\dot{\beta}} W^I\cr
\epsilon^{\dot{\alpha}} &=& \Gamma^{I\dot{\alpha}\beta} \epsilon_I U_{\beta}
\eea
Here $X^I$ denote the eight scalar fields in $8_v$, $\psi_{\alpha}$ denotes the $8_s$-spinors and $\epsilon^{\dot{\alpha}}$ the supersymmetry parameter in $8_c$. For single index quantities we use the convention that rising and lowering indices correspond to complex conjugation. For multiple index quantities we have also to specify an ordering prescription when rising and lowering the indices under complex conjugation. An example of this is (\ref{ordering}). It is true that we can stick to a basis where all entries are real in the gamma matrices and the Majorana spinors, but things get more transparent if we work in a general basis. 

We have the following identities
\bea
\Gamma^{I\dot{\alpha}\beta} U_{\beta} U^{\gamma} \Gamma_{I\gamma\dot{\delta}} &=& \delta^{\dot{\alpha}}_{\dot{\delta}}\label{fierz1}
\eea
\bea
\Gamma_{J\alpha\dot{\gamma}} V^{\dot{\gamma}} V_{\dot{\beta}} \Gamma^{I\dot{\beta}\alpha}&=& \delta_J^I
\eea
\bea
W_I \Gamma^{I\dot{\beta}\alpha}\Gamma_{J\alpha\dot{\gamma}}W^J &=& \delta_{\dot{\gamma}}^{\dot{\beta}}
\eea
To prove the first of these identities we note the Fierz rearrangement
\bea
U U^{\dag} &=& \frac{1}{16} \Big( U^{\dag} U + \frac{1}{12} U^{\dag} \Gamma^{IJKL} U \Gamma_{IJKL} \Big) (1+\Gamma)
\eea
together with the gamma matrix identity 
\bea
\Gamma_{I} \Gamma_{JKLM} \Gamma^{I} &=& 0
\eea
To prove the second of these identities we use the Clifford algebra. The third identity follows by trace properties of the gamma matrices. 

Armed with these identities we can map any contraction to its trial contraction, for example
\bea
X^{\alpha} Y_{\alpha} &=& (\Gamma^{I\dot{\beta}\alpha}X_I V_{\dot{\beta}}) (\Gamma_{J\alpha\dot{\gamma}} Y^J V^{\dot{\gamma}})\cr
&=& \Gamma_{J\alpha\dot{\gamma}} V^{\dot{\gamma}} V_{\dot{\beta}} \Gamma^{I\dot{\beta}\alpha} X_I Y^J\cr
&=& X_I Y^I
\eea

\section{Relating BLG with ABJM}
In string theory we are familiar with that two different theories can be unified if we move one dimension higher. In this section we will recall how BLG and ABJM theories, which are in general unrelated in two auxiliary dimensions (the only exception being when the gauge group is $SU(2)\times SU(2)$), get unified in three auxiliary dimensions where we can use a star-3-product \cite{Gustavsson:2010ep}. 
  
In ABJM theory we have a three-bracket defined as \cite{Bagger:2008se}
\bea
[T^a,T^b;T^c] &=& T^a T_c T^b - T^b T_c T^a
\eea
The generators $T^a$ are usually taken to be $N\times N$ matrices and the gauge group $U(N)\times U(N)$. We can also use star-products of functions. These functions live on a two-manifold. It is well-known how the star-product is mapped isomorphically to matrix multiplication when we have either a two-sphere or a two-torus, and in this paper we will only consider the two-torus. The idea is to describe a manifold by the algebra of functions. While it is true that the functions may live on a smooth and classical two-torus, if we only have a finite set of functions we can not probe the smooth structure of the two-torus. Instead the torus will appear like a fuzzy or noncommutative manifold where the coordinates do not quite commute. The finite set of functions correspond to the finite rank of the corresponding matrices. By taking $N\rightarrow \infty$ we obtain the smooth manifold. 

Let us now assume that we are given a three-manifold $M_3$ and some three-algebra generators $T^a$ which are functions on $M_3$. We denote by $T_a = (T^a)^*$ the complex conjugated elements. In general the star-3-product defined as
\bea
T^a * T_c * T^b &=& \exp\left\{\frac{\hbar}{2}\sqrt{g}\epsilon^{\alpha\beta\gamma}\partial_{\alpha}\partial'_{\beta}\partial''_{\gamma}\right\} T^a(\sigma) T^b(\sigma') T_c(\sigma'')|_{\sigma = \sigma' = \sigma''}
\eea
is not associative. We will now assume that $M_3$ is a circle bundle and let 
\bea
\sigma^{\alpha} &=& (\sigma^a,\psi = \sigma^3)
\eea
be coordinates on $M_3$, $\sigma^a$ (for $a=1,2$) be coordinates on a two-dimensional base-manifold, and $\psi$ be a coordinate on the fiber. It is necessary that we restrict ourselves to functions on $M_3$ which are on the form
\ben
T^a &=& e^{i\psi} \t T^a(\sigma^a)\label{trivial}
\een
in order for the star-3-product to become associative. We denote by $g_{\alpha\beta}$ the metric on the three-manifold, and by $G_{ab}$ the metric on the base manifold. We define the totally antisymmetric tensors like $\epsilon_{123} = 1$ and rise all indices by the inverse metrics, so for example $g \epsilon^{123} = 1 = G \epsilon^{12}$. We define a star-2-product as
\bea
T^a * T^b &=& \exp\left\{\frac{i\cepsilon}{2}\sqrt{G}\epsilon^{ab}\partial_a \partial_b'\right\} T^a(\sigma) T^b(\sigma')
\eea
The relation between $\cepsilon$ and $\hbar$ reads
\bea
\cepsilon &=& \hbar \sqrt{\frac{G}{g}}
\eea
The associated star-commutator is given by
\bea
[f,g] &=& i\epsilon \{f,g\} + \O(\epsilon^2)\cr
\{f,g\} &=& \sqrt{G}\epsilon^{ab} \partial_a f \partial_b g
\eea
The star-3-product we use is not a genuine star-3-product since we restrict ourselves to functions that are essentially living on the base manifold, all having the same rather trivial dependence on the fiber according to Eq (\ref{trivial}). Moreover, just as one should expect of such a star-3-product, it can be expressed as a composition of two consecutive star-2-products, 
\bea
T^a * T_c * T^b &=& (T^a * T_c) * T^b
\eea
We define a totally antisymmetric three-bracket as
\bea
[T^a,T^b,T_c] &=& T^a * T_c * T^b - T^b * T_c * T^a
\eea
We will refer to this bracket as the star-3-commutator. To first order in $\hbar$ it is given by 
\bea
[T^a,T^b,T_c] &=& \hbar \{T^a,T^b,T_c\}
\eea
In an appendix in \cite{Gustavsson:2010nc} it is shown that the star-3-commutator is totally antisymmetric to all orders.

In BLG theory we need a totally antisymmetric three-bracket. As we have shown, we may use the star-3-commutator on a certain two-dimensional subset of functions on a circle-bundle over a two-manifold. By triality of $SO(8)$ we may always assume that the field content of BLG theory consists of eight scalars $X_{\alpha}$ in $8_s$ and eight fermions $\psi^{\dot{\alpha}}$ in $8_c$. The sextic potential is given by
\bea
V &=& \frac{1}{12}|[X^{\alpha},X^{\beta},X^{\gamma}]|^2
\eea
To connect with ABJM theory we decompose the scalar fields as
\bea
X^{\alpha} &=& \(\begin{array}{c}
Z^A(x^a) e^{i\psi}\\
Z_A(x^a) e^{-i\psi}
\end{array}\)
\eea
This decomposition breaks $SO(8)$ down to $SO(6)$ whereof $Z^A$ is a Weyl spinor. Though the more common way of expressing the same thing is as the defining representation of $SU(4) \simeq SO(6)$. By utilizing the total antisymmetry of the three-bracket, we expand out the sextic potential as
\bea
V &=& \frac{1}{12}\(2|[Z^A,Z^B,Z^C]|^2+6|[Z^A,Z^B,Z_C]|^2\)
\eea
By using the fundamental identity we derive the identity
\bea
|[Z^A,Z^B,Z^C]|^2 &=& |[Z^A,Z^B,Z_C]|^2 - 2 |[Z^A,Z^B,Z_B]|^2
\eea
and we find the ABJM sextic potential
\bea
V &=& \frac{2}{3}\(|[Z^A,Z^B,Z_C]|^2 - \frac{1}{2} |[Z^A,Z^B,Z_B]|^2\)
\eea
We may restore the ABJM three-bracket and we get
\bea
V &=& \frac{2}{3}\(|[Z^A,Z^B;Z^C]|^2 - \frac{1}{2} |[Z^A,Z^B;Z^B]|^2\)
\eea
Our first observation now is that only $Z^A$ occurs in this final expression, and no $Z_A$. This means that only $T^a$ three-algebra generators arise in this expression, and no $T_a$. By taking $T^a = e^{i\psi}\t T^a$, the ABJM three-bracket reduces as
\bea
[T^a,T^b;T^c] &=& T_c [T^a,T^b] + [T^a,T_c]T^b - [T^b,T_c]T^a
\eea
In the RHS we have usual star-product multiplications (star-2-products). Mapping these to matrix multiplications, we make contact with ABJM theory as it was originally formulated. 

We conclude that the bosonic part of the ABJM Lagrangian can be expressed as
\bea
\L &=& \frac{KN}{2\pi} \(\L_{kin} + \L_{CS} + \L_{pot}\)
\eea
where
\bea
\L_{kin} &=& -\frac{1}{2}|D_{\mu}X^{\alpha}|^2\cr
\L_{CS} &=& \frac{1}{2}\epsilon^{\mu\nu\lambda}\Big(\<T^b,[T^c,T^d;T^a]\>A_{\mu}{}^c{}_b \partial_{\nu} A_{\lambda}{}^d{}_a\cr
&&-\frac{2}{3}\<[T^a,T^c;T^d],[T^f,T^b;T^e]\>A_{\mu}{}^b{}_a A_{\nu}{}^d{}_c A_{\lambda}{}^f{}_e\Big)\cr
\L_{pot} &=& -\frac{1}{12} |[X^{\alpha},X^{\beta};X^{\gamma}]|^2
\eea

\section{The M5 brane solution}
We will now review the solution that we presented in the introduction, closely following the original work \cite{Terashima:2009fy}. We decompose the ABJM scalar fields as
\bea
Z^A &=& \(\begin{array}{c}
Z^a\\
Z_{\dot{a}}
\end{array}\)
\eea
corresponding to $SU(4) \rightarrow SU(2)\times SU(2)$, and make the following ansatz for these components,
\ben
Z^a &=& x^a e^{i\psi}\cr
Z_{\dot{a}} &=& 0\label{threemanifold}
\een
Here
\bea
0 \leq \psi \leq \frac{2\pi}{K}
\eea
and $x^a$ are real. We assume that the metric on the base manifold is given by 
\bea
ds^2 &=& \delta_{ab}dx^a dx^b
\eea
and we define the metric tensor $G_{ab} = \delta_{ab}$ and its determinant is $G = 1$. 

By solving the equation of motion we will now determine the metric on the three-manifold $M_3$, whose base manifold is the two-dimensional plane described above. We make the ansatz
\bea
ds^2 &=& \delta_{ab} dx^a dx^b + f(x^a) d\psi^2
\eea
If we let $x^{\alpha} = (x^a,\psi)$ denote the coordinates on $M_3$ the ansatz for the metric tensor reads
\bea
g_{\alpha\beta} &=& \(\begin{array}{cc}
\delta_{ab} & 0\\
0 & f
\end{array}\)
\eea
and its determinant is $g = f$. Here $f$ is a function that is to be determined by solving the static equation of motion of ABJM theory. 

Inserting our ansatz into the sextic potential (ignoring any overall factor)
\bea
|\[Z^A,Z^B;Z^C\]|^2 - \frac{1}{2}|\[Z^A,Z^B;Z^B\]|^2
\eea
it reduces to
\bea
|\[x^1,(x^2)^2\]|^2 + |\[x^2,(x^1)^2\]|^2
\eea
We now vary $x^1$. This gives us the equation of motion
\bea
\[(x^2)^2,\[x^1,(x^2)^2\]\] + \(x^1,\[x^2,\[(x^1)^2,x^2\]\]\) &=& 0
\eea
and a corresponding equation obtained by exchanging indices $1$ and $2$. 

The noncommutativity parameter that sits in the star-2-product becomes
\bea
\cepsilon &=& \frac{\hbar}{\sqrt{f}}
\eea
To lowest order in this parameter, the star-2-(anti-)commutator is given by
\bea
[x^1,x^2] &=& \frac{i\hbar}{\sqrt{f}} \{x^1,x^2\}\cr
(x^1,x^2) &=& 2 x^1 x^2
\eea
where the Poisson bracket is given by 
\bea
\{x^1,x^2\} &=& 1
\eea
We now get the equation of motion as
\bea
\((x^1)^2 + (x^2)^2\) [x^2,[x^1,x^2]] - x^1 [x^1,x^2]^2 &=& 0
\eea
By further noting that 
\bea
[x^2,\bullet] &=& -\frac{i \hbar}{\sqrt{f}} \frac{\partial}{\partial x^1}
\eea
we can express the equation of motion as
\bea
\((x^1)^2 + (x^2)^2\) \frac{\partial}{\partial x^1} \(\frac{1}{\sqrt{f}}\) + x^1 \frac{1}{\sqrt{f}} &=& 0
\eea
We have a similar equation from varying $Z^2$ which is obtained by exchanging indices $1$ and $2$. The solution to these two equations is given by 
\bea
f &=& C\((x^1)^2 + (x^2)^2\)
\eea
and thus we deduce that any three-manifold on with the metric
\bea
ds^2 &=& (dx^1)^2 + (dx^2)^2 + C\((x^1)^2 + (x^2)^2\) d\psi^2
\eea
for any constant $C$, solves the ABJM equation of motion. But we can absorb this constant into $\psi$ by rescaling $\psi$. Hence we can always assume that the metric is given by 
\ben
ds^2 &=& (dx^1)^2 + (dx^2)^2 + r^2 d\psi^2\label{metric}
\een
where we define
\bea
r^2 &=& (x^1)^2 + (x^2)^2
\eea
This now, is the metric that is induced from the flat metric on $\mb{C}^4/{\mb{Z}_K}$,
\bea
ds^2 &=& dZ^a dZ_a + dZ_{\dot{a}} dZ^{\dot{a}}
\eea 
We now also see that we shall let $0\leq \psi \leq \frac{2\pi}{K}$ if we consider the orbifold $\mb{C}^4/{\mb{Z}_K}$. This M5 brane solution is valid for any integer value on $K$. 

If we parameterize the base two-manifold by the coordinates $(x^1,x^2)$ in which the metric is a given above, then the non-commutativity parameter in the star-2-product is given by
\bea
\cepsilon &=& \frac{\hbar}{r}
\eea
and is clearly non-constant. 

The star-2-product is still associative and the Jacobi identity is still satisfied, even for a non-constant non-commutativity parameter. To check this, we just have to notice that $\epsilon^{a[b} \epsilon^{cd]} = 0$, which is true in two dimensions since we antisymmetrize over three indices. 

The three-manifold $M_3$ can alternatively be expressed as an embedded surface in $\mb{C}^2$ as
\bea
T^1 T_2 - T^2 T_1 &=& 0
\eea
where $T_a = (T^a)^*$. The solution obeys the three-algebra
\bea
\{T^1,T^2,T_2\} &=& -2i T^2\cr
\{T^2,T^1,T_1\} &=& 2i T^1
\eea
We can bring this into the standard form of $SO(4)$ three-algebra by defining
\bea
S^1 &=& T^1 + i T^2\cr
S^2 &=& T^1 - i T^2
\eea
Then we find 
\bea
\{S^a,S^b,S_c\} &=& -8 \delta^{ab}_{cd} S^d
\eea
which is the standard $SO(4)$ three-algebra expressed in a complex basis. We have difficulties finding finite-dimensional matrix representations of this algebra which also satisfy the condition 
\bea
S^1 S_1 - S^2 S_2 &=& 0
\eea
which describes the embedded three-manifold $M_3$ in these new coordinates. 

Finally we can express $M_3$ as the embedding 
\bea
y &=& 0
\eea
by choosing the coordinates as
\bea
z^a &=& \( x^a + i y \frac{\epsilon^{ab} x_b}{r} \) e^{i\psi}
\eea
In polar coordinates
\bea
r e^{i\varphi} &=& x^1 + i x^2
\eea
we have the coordinate transformation
\bea
s^1 &=& (r+y)e^{i(\psi+\varphi)}\cr
s^2 &=& (r-y)e^{i(\psi-\varphi)}
\eea
where thus $s^a|_{y=0} = S^1$, and the metric becomes
\bea
ds^2 &=& \frac{1}{2}\(|ds^1|^2 + |ds^2|^2\)\cr
&=& dr^2 + dy^2 + (r^2 + y^2) (d\psi^2 + d\varphi^2) + 2 ry d\psi d\varphi
\eea
Transformed back to cartesian coordinates on the 2-plane, this reads
\bea
ds^2 &=& \delta_{ab} dx^a dx^b + r^2 d\psi^2 + dy^2 + \O(y)
\eea
We see that $y$ is a normal direction to $M_3$ 
\bea
T^a &=& z^a|_{y=0}
\eea
and there is no off-diagonal metric components along the $y$-direction,
\bea
g_{y a} &=& 0\cr
g_{y \psi} &=& 0
\eea
when we confine ourselves to $M_3$.

\section{Local quantization of the solution}
We have not managed to quantize $M_3$. But also, we can not quantize $\mb{R}^2$ since this a non-compact space. What we can do is to consider a local two-torus somewhere on $\mb{R}^2$, far away from the curvature singularity at the origin. To this end we will express the vacuum solution as 
\ben
T^a &=& \(v^a + R\sigma^a\) e^{i\psi}\label{vev}
\een
and we will assume that $v^a>>R$ and let $0\leq \sigma^a\leq 2\pi$ parametrize the local two-torus. Here $R$ is a length scale of this two-torus. The metric is 
\bea
ds^2 &=& R^2\delta_{ab} d\sigma^a d\sigma^b + v^2\(1 + \O\(\frac{R}{v}\)\) d\psi^2
\eea
where $v = \sqrt{\delta_{ab} v^a v^b}$. We will treat $\frac{R}{v}$ as an expansion parameter, which will enable us to perform a systematic fluctuation analysis to obtain the D4 brane Lagrangian. We have the square root determinants of the metrics
\bea
\sqrt{G} &=& R^2\\
\sqrt{g} &=& R^2 v \(1 + \O\(\frac{R}{v}\)\)
\eea
The relation between the two-dimensional and three-dimensional non-commutativity parameters then reads  
\bea
\cepsilon &=& \frac{\hbar}{v}\(1 + \O\(\frac{R}{v}\)\)
\eea
and the quantization condition for the two-dimensional non-commutativity parameter can be inferred from the fuzzy two-torus structure\footnote{A definition of the fuzzy two-torus in the conventions of this paper is found in \cite{Gustavsson:2010nc}. A review paper of fuzzy manifolds is \cite{Taylor:2001vb} where also many references can be found to fuzzy Riemann manifolds.}
\bea
\cepsilon &=& \frac{2\pi R^2}{N}
\eea
where $N$ is the size of the corresponding matrix realization of the fuzzy two-torus. 

It is important to note that the background is the two-torus $T^2$ and not the point $Z^a = v^a e^{i\psi}$. But this may at first sight seem confusing since then $Z^a = Z^a(\sigma)$ are functions rather than taking specific values as is the usual situation when one gives a vacuum expectation value to a scalar field. But here the scalar fields defined by Eq (\ref{vev}) on the whole $T^2$ is really our vacuum expectation value. The intuitive picture is that the vacuum expectation value is an infinite-dimensional diagonal matrix whose eigenvalues are the different points on $T^2$. At each point we have an M2 brane with worldvolume $\mb{R}^{1,2}$, and hence the collection of all these M2 branes give us a D4 brane whose world-volume is $\mb{R}^{1,2} \times T^2$. This picture is somewhat intuitive though. In reality we have a finite set of M2 branes and a fuzzy $T^2$. Since it is fuzzy, we have a Heisenberg type uncertainty forbidding us to pack the M2 branes too dense, thus there is room only for a finite number of M2 branes on the fuzzy $T^2$.

Since $\sigma^a \sim \sigma^a + 2\pi$, we also find that the $T^a$ are compact. Accordingly we shall also take the ABJM scalar fields to be compact,
\bea
Z^a &\sim & Z^a + 2\pi R
\eea
We will for the most part of this paper consider only small fluctuations around (\ref{vev}), much smaller than the size of $T^2$, and so the fact that these scalar fields are compact can be largely ignored.

\section{The Higgs mechanism}
To study the Higgs mechanism, the first thing we need to do is to consider the covariant derivative acting on a scalar field that we will eventually give a vacuum expectation value. The covariant derivative is given by 
\bea
D_{\mu} Z^A &=& \partial_{\mu} Z^A + [Z^A,T^a;T^b] A_{\mu}{}^b{}_a
\eea
We define
\bea
A^- &=& \frac{i}{2} [T^a,T_b] A_{\mu}{}^b{}_a\cr
A^+ &=& \frac{i}{2} (T^a,T_b) A_{\mu}{}^b{}_a
\eea
where we use round brackets for the anticommutator. We now get
\bea
D_{\mu} Z^A &=& \partial_{\mu} Z^A + i[A^+_{\mu},Z^A] + i (Z^A,A^-_{\mu})
\eea

We give a Higgs vacuum expectation value to the scalar fields,
\bea
Z^A &=& T^A + Y^A
\eea
Here $T^A$ is the vacuum expectation value, and $Y^A$ are the fluctuations. We then expand the covariant derivative, and find
\bea
D_{\mu} Z^A &=& D_{\mu} Y^A + i(T^A,A_{\mu}^-) + i[A_{\mu}^+,T^A]
\eea
where we define
\bea
D_{\mu} Y^A &=& \partial_{\mu} Y^A + i [A_{\mu}^+,Y^A]
\eea
where, if we can neglect $i(Y^A,A_{\mu}^-)$, we have isolated the term that involves $A^-_{\mu}$ from all the rest. Eventually we will discover that $A^-_{\mu}$ enters the Lagrangian only algebraically and can be integrated out. 

We define the induced metric tensor as
\bea
G_{ab} &=& \partial_{(a} T^A \partial_{b)} T_A\cr
G_{IJ} &=& \partial_{(I} T^A \partial_{J)} T_A
\eea

\section{Fluctuation analysis}
We will now expand the ABJM Lagrangian about the vacuum solution where we keep only the leading order terms in the expansion parameter $\frac{R}{v}$.

\subsection{The kinetic term}
We expand out the kinetic term
\bea
\L_{kin} &=& - \<D_{\mu}Z^A,D_{\mu}Z^A\>
\eea
around the Higgs vacuum expection value. We define
\bea
Y^A &=& Y^a \partial_a T^A + Y^I \partial_I T^A\cr
Y^a &=& \lambda \sqrt{G} \epsilon^{ab} A_b\cr
Y^I &=& \lambda \phi^I
\eea
and define
\bea
A_{\mu}^+ &=& -\frac{\lambda}{\cepsilon} A_{\mu}
\eea
We then get
\bea
D_{\mu} Z^A &=& \lambda \(\sqrt{G} \epsilon^{ab} F_{\mu b} \partial_a T^A + D_{\mu} \phi^I \partial_I T^A\) + 2i T^A A_{\mu}^-
\eea
where
\bea
F_{\mu a} &=& \partial_{\mu} A_a - \partial_a A_{\mu} - \frac{i\lambda}{\cepsilon} [A_{\mu},A_a]\cr
D_{\mu} \phi^I &=& \partial_{\mu} \phi^I - \frac{i\lambda}{\cepsilon} [A_{\mu},\phi^I]
\eea
Then we get
\bea
\L_{kin} &=& -\lambda^2 \(F_{\mu a} F^{\mu a} + G_{IJ} D_{\mu} \phi^I D^{\mu} \phi^J\) - 2 T^{\alpha} T_{\alpha} A_{\mu}^- A^{-\mu}
\eea

\subsection{The Chern-Simons term}
The Chern-Simons term becomes
\bea
\L_{CS} &=& \epsilon^{\mu\nu\lambda} \<A^-_{\mu}F^+_{\nu\lambda} - \frac{2i}{3} A^-_{\mu} A^-_{\nu} A^-_{\lambda}\>
\eea
where
\bea
F_{\mu\nu}^+ &=& \partial_{\mu}A_{\nu}^+ - \partial_{\nu}A_{\mu}^+ - i [A_{\mu}^+,A_{\nu}^+]
\eea
Alternatively
\bea
\L_{CS} &=& -\frac{\lambda}{\cepsilon} \epsilon^{\mu\nu\lambda} \<A^-_{\mu}F_{\nu\lambda}\>
\eea
where
\bea
F_{\mu\nu} &=& \partial_{\mu}A_{\nu}-\partial_{\nu}A_{\mu} - \frac{i\lambda}{\cepsilon}[A_{\mu},A_{\nu}]
\eea

\subsection{The sextic potential}
\subsubsection{Quadratic order}
From a technical point of view, the ABJM sextic potential is highly complicated to expand. It here advantegous to make use of the BLG formulation instead. It is useful to define a sign
\bea
s(\alpha) &=& \pm 1
\eea
according to whether $\aalpha = {}^A$ or $\aalpha = {}_A$. That is, $s({}^A) =1$ and $s({}_A) = -1$. Then we have
\bea
\partial_{\psi}X^{\alpha} &=& i s(\alpha) X^{\alpha}
\eea
and we have the following useful result,
\bea
s(\alpha) \partial_m T^{\alpha} \partial_n T_{\alpha} &=& 0
\eea
Instead of expanding the ABJM sextic potential, we may now instead expand the equivalent BLG sextic potential. At the moment we will be ignorant about the overall factor, which we will determined later. At quadratic order in $Y_{\alpha}$ we then consider the following terms, 
\bea
\hbar^{-2} V &=& \frac{1}{12}\{X^{\alpha},X^{\beta},X^{\gamma}\}\{X_{\alpha},X_{\beta},X_{\gamma}\} \cr&=& \frac{1}{4}\{T^{\alpha},T^{\beta},Y^{\gamma}\}\{T_{\alpha},T_{\beta},Y_{\gamma}\}+ \frac{1}{2}\{T^{\alpha},T^{\beta},Y^{\gamma}\}\{T_{\alpha},Y_{\beta},T_{\gamma}\}+ \frac{1}{2}\{T^{\alpha},T^{\beta},T^{\gamma}\}\{T_{\alpha},Y_{\beta},Y_{\gamma}\}
\eea
We write this out explicitly as
\bea
\hbar^{-2} V &=& 2g^{\gamma\gamma'} \partial_{\gamma}Y^{\ggamma} \partial_{\gamma'}Y_{\ggamma} \cr
&&+ \(g^{\beta\beta'}g^{\gamma\gamma'}-g^{\beta\gamma'}g^{\gamma\beta'}\) \(\partial_{\gamma'}T_{\ggamma}\partial_{\gamma}Y^{\ggamma} + \partial_{\gamma}T_{\ggamma}\partial_{\gamma'}Y^{\ggamma}\) \partial_{\beta} T^{\bbeta} \partial_{\beta'}Y_{\bbeta}
\eea
where we now start to underline the $SO(8)$ indices $\aalpha$ in order to distinguish them from the indices $\alpha$ on $M_3$. We now need the derivatives
\bea
\partial_{\psi} Y^{\aalpha} &=& i s(\aalpha) Y^{\aalpha}\cr
\partial_m Y^{\aalpha} &=& \partial_m Y^n \partial_n T^{\aalpha} + \partial_m Y^I \partial_I T^{\aalpha} + Y^y \partial_m \partial_y T^{\aalpha}
\eea
We will neglect the last term which is
\bea
Y^y \partial_m \partial_y T^{\aalpha} &=& \O\(\frac{R}{v}\)
\eea
We split the indices as $\aalpha = (a,\psi)$ and accordingly we split $\hbar^{-2} V = V_{(I)} + V_{(II)} + V_{(III)}$ where
\bea
V_{(I)}&=&2 G^{ab} \partial_a Y^{\ggamma} \partial_{b} Y_{\ggamma} \cr
&&+\(G^{ab}G^{cd} - G^{ad}G^{cb}\) \(\partial_{d}T_{\ggamma}\partial_c Y^{\ggamma} + \partial_c T_{\ggamma} \partial_{d} Y^{\ggamma}\)  \partial_a T^{\bbeta} \partial_{b}Y_{\bbeta}\cr
V_{(II)} &=& 2g^{\psi\psi} Y^{\ggamma} Y_{\ggamma}\cr
V_{(III)} &=& 4 G^{ab} g^{\psi\psi} T^{\aalpha} Y_{\aalpha} \partial_a T^{\bbeta} \partial_b Y_{\bbeta}
\eea
We have also the mixed terms, which can be gathered into 
\bea
G^{ab} g^{\psi\psi} \partial_a (T^{\aalpha} Y_{\aalpha}) \partial_b (T^{\bbeta} Y_{\bbeta})s(\aalpha) s(\bbeta) 
\eea
and this vanishes by 
\bea
s(\aalpha) T^{\aalpha} Y_{\aalpha} &=& 0
\eea
it being understood that this is to be summed over $\aalpha$. For the remaining pieces we find
\bea
V_{(III)} &=& \frac{16 \lambda^2}{v^2} G \epsilon^{ab} \epsilon^{cd} v_a A_b \partial_c A_d
\eea
\bea
V_{(II)} &=& \frac{4\lambda^2}{v^2} \(G^{ab} A_a A_b + G_{IJ} \phi^I \phi^J\)
\eea
\bea
V_{(I)} &=& 2\lambda^2 f_{ab} f^{ab} + 4\lambda^2 \partial_a \phi^I \partial^a \phi_I + \O\(\frac{R}{v}\)
\eea

We now see that in order to get the kinetic term on the form
\bea
-\lambda^2 G_{IJ} \(\eta^{\mu\nu} D_{\mu} \phi^I D_{\nu} \phi^J + G^{ab} D_a \phi^I D_b \phi^J\)
\eea
we must rescale the scalar fields in BLG theory. A rescaling 
\bea
X^{\aalpha} &\rightarrow & \mu X^{\aalpha}
\eea
yields 
\bea
\mu^2\(-\frac{1}{2}\<D_{\mu} X^{\aalpha},D^{\mu} X^{\aalpha}\> - \frac{\mu^4}{12} \<[X^{\aalpha},X^{\bbeta};X^{\ggamma}],[X^{\aalpha},X^{\bbeta};X^{\ggamma}]\>\)
\eea
Here we shall take 
\bea
\mu^2 &=& \frac{1}{2 \hbar}
\eea
Then the BLG Lagrangian reads
\bea
&&\frac{1}{2 \hbar} \(-\frac{1}{2}\<D_{\mu} X^{\aalpha},D^{\mu} X^{\aalpha}\> - \frac{1}{48\hbar^2} \<[X^{\aalpha},X^{\bbeta};X^{\ggamma}],[X^{\aalpha},X^{\bbeta};X^{\ggamma}]\>\)\cr
&&+\frac{i\lambda}{\cepsilon} \epsilon^{\mu\nu\lambda} \<A_{\mu}^- F_{\nu\lambda}\>
\eea
and we wish to integrate out $A_{\mu}^-$. We collect terms that contain $A_{\mu}^-$,
\bea
&&\frac{1}{2\hbar} 2 T^{\aalpha} T_{\aalpha} A_{\mu}^- A^{-\mu} + \frac{i\lambda}{\cepsilon} \epsilon^{\mu\nu\lambda}A_{\mu}^- F_{\nu\lambda}\cr
&\rightarrow & \(\frac{\lambda}{\cepsilon}\)^2\frac{\hbar}{2 T^{\aalpha} T_{\aalpha}} F_{\mu\nu} F^{\mu\nu}\(1 + \O\(\frac{R}{v}\)\)
\eea
We now note that
\bea
T^{\aalpha} T_{\aalpha} = 2T^A T_A = 2 v^2\(1 + \O\(\frac{R}{v}\)\)
\eea
and by re-instating the overall factor and the explicit realization of the inner product as $\frac{KN}{2\pi}\int \frac{d^2\sigma}{4\pi^2}$, we get
\bea
\int d^2 \sigma \sqrt{G} \frac{1}{4\pi^2 R^2} \frac{KN}{2\pi} \(\frac{\lambda}{\cepsilon}\)^2 \frac{2\pi R^2 v}{2N 2R^2 v^2} F_{\mu\nu} F^{\mu\nu}\(1 + \O\(\frac{R}{v}\)\)
\eea
where we have rewritten the measure in a covariant form and used the fact that $\sqrt{G} = R^2$. Simplifying these factors, we end up with
\bea
\int d^2 \sigma \sqrt{G} \(\frac{\lambda}{\cepsilon}\)^2 \frac{K}{4\pi^2 v} \frac{1}{4} F_{\mu\nu} F^{\mu\nu} \(1 + \O\(\frac{R}{v}\)\)
\eea
and we can already here read off the Yang-Mills coupling constant as
\bea
g_{YM}^2 &=& 4\pi^2 \frac{v}{K} \(1 + \O\(\frac{R}{v}\)\)
\eea

Once we now have obtained the correct normalization of the sextic potential, we can now start computing it to various orders in the fluctuation fields. 

\subsubsection{Zeroth order}
At zeroth order we have
\bea
\<[T^{\aalpha},T^{\bbeta};T^{\ggamma}],[T^{\aalpha},T^{\bbeta};T^{\ggamma}]\> &=& \hbar^2 g \epsilon^{\alpha\beta\gamma}\epsilon^{\alpha'\beta'\gamma'} 8 g_{\alpha\alpha'} g_{\beta\beta'} g_{\gamma\gamma'}\<1\>\cr
&=& 48 \hbar^2 \<1\>
\eea
Here 
\bea
\<1\> &=& \int \frac{d^2 \sigma}{(2\pi)^2} 1 = 1
\eea
Now
\bea
\L_{pot} &=& 
-\frac{KN}{2\pi 96\hbar^4} \<[T^{\aalpha},T^{\bbeta};T^{\ggamma}],[T^{\aalpha},T^{\bbeta};T^{\ggamma}]\>\cr
&=& - \frac{KN}{4\pi\hbar^2}\cr
&=& - \frac{KN^3}{16\pi^3 R^4 v^2}
\eea

\subsubsection{Linear order}
At linear order we have $6$ identitcal terms,
\bea
&&6\<[T^{\aalpha},T^{\bbeta};T^{\ggamma}],[T^{\aalpha},T^{\bbeta};Y^{\ggamma}]\>\cr
&=& 6\hbar^2 g \epsilon^{\alpha\beta\gamma}\epsilon^{\alpha'\beta'\gamma'} \partial_{\alpha} T^{\aalpha} \partial_{\beta} T^{\bbeta} \partial_{\gamma} T^{\ggamma} \partial_{\alpha'} T_{\aalpha} \partial_{\beta} T_{\bbeta} \partial_{\gamma} Y_{\ggamma} \cr
&=& 48\hbar^2 \lambda \sqrt{G} \epsilon^{ab} f_{ab}
\eea
Including the correct overall normalization and the combinatorical factor of $6$, we have
\bea
\L_{pot} &=& - \frac{\lambda}{2 \hbar^2} \sqrt{G} \epsilon^{ab} f_{ab}
\eea
Now we must also find the non-Abelian completion of this term in the higher order terms.

\subsubsection{Quadratic order}
The terms that we previously overlooked at quadratic order are given by (there are $6$ terms of this type)
\bea
&&\<[T^{\aalpha},T^{\bbeta},T^{\ggamma}],[T^{\aalpha},Y^{\bbeta};Y^{\ggamma}]\>\cr
&=& -8i\cepsilon v^2 \lambda^2 \sqrt{G} \epsilon^{ab} [A_a,A_b]
\eea
This combines with the linear order term into 
\bea
8\hbar^2 \lambda \sqrt{G} \epsilon^{ab} F_{ab}
\eea
where
\bea
F_{ab} &=& \partial_a A_b - \partial_b A_a - \frac{i\lambda}{\cepsilon}[A_a,A_b].
\eea

We have from earlier computation
\bea
\<[XXX],[XXX]\>|_{quadratic} &=& 24 \hbar^2 \lambda^2 f_{ab} f^{ab}
\eea
which combines with the linear and zeroth order terms into
\bea
&& 48 \hbar^2 \(1 + \lambda \sqrt{G}\epsilon^{ab} f_{ab} + \frac{1}{2} \lambda^2 f^{ab} f_{ab}\)\cr
&=& 24 \hbar^2 \lambda^2 \(f_{ab} + \lambda^{-1}\sqrt{G}\epsilon_{ab}\)\(f^{ab} + \lambda^{-1}\sqrt{G}\epsilon^{ab}\)
\eea
Normalizing and rewriting the trace form as
\bea
\<...\> &=& \frac{1}{4\pi^2 R^2} \int d^2 \sigma \sqrt{G}
\eea
we have
\bea
-\frac{KN}{2.96\pi \hbar^3}\frac{1}{4\pi^2 R^2} \int d^2\sigma \sqrt{G} [XXX][XXX] &=& -\(\frac{\lambda}{\cepsilon}\)^2 \frac{K}{16\pi^2 v} \int d^2 \sigma \sqrt{G} \t f_{ab} \t f^{ab}
\eea
where 
\bea
\t f_{ab} &=& f_{ab} + \lambda^{-1} \sqrt{G}\epsilon_{ab}
\eea

\subsubsection{Cubic interactions}
First we note that there are no contributions on the form
\bea
\<[T^{\aalpha},T^{\bbeta},T_{\ggamma}]_{\A},[Y^{\aalpha},Y^{\bbeta};Y^{\ggamma}]_{\B}\>
\eea
Using the three-algebra satisfied by the $T^{\aalpha}$, we can write this as a sum of terms each of the form
\bea
\<T^{\beta},[Y^{\alpha},Y^{\beta};Y^{\alpha}]\>
\eea
Now this vanishes due to $T^{\aalpha} Y_{\aalpha} = 0$. 

We next note that
\bea
[X^{\aalpha},X^{\bbeta};X^{\ggamma}] &=& X_{\ggamma} [X^{\aalpha},X^{\bbeta}]_{s(\gamma)} + [X^{\aalpha},X^{\bbeta}]_{s(\beta)} X^{\bbeta} - [X^{\bbeta},X_{\ggamma}]_{s(\aalpha)} X^{\aalpha}
\eea
where the subscripts indicate which sign to use for the non-commutativity parameter, that is $s(\aalpha)\cepsilon$, in the star-commutator. We keep the following terms
\bea
[T^{\aalpha},T^{\bbeta};Y^{\ggamma}] &=& [T^{\aalpha},Y_{\ggamma}]_{s(\bbeta)} T^{\bbeta} - [T^{\bbeta},Y_{\ggamma}]_{s(\aalpha)} T^{\aalpha}\cr
[T^{\aalpha},Y^{\bbeta};Y^{\ggamma}] &=& -[Y^{\bbeta},Y_{\ggamma}]_{s(\alpha)} T^{\aalpha}
\eea
and we get
\bea
\<[T^{\aalpha},T^{\bbeta};Y^{\ggamma}],[T^{\aalpha},Y^{\bbeta};Y^{\ggamma}]\> &=& 8i v^2 \cepsilon \lambda^3 G^{ab} \(G^{cd} \<\partial_a A_c,[A_b,A_d]\> + G_{IJ} \<\partial_a\phi^I,[A_b,\phi^J]\>\)
\eea

\subsubsection{Quartic interaction}
We only get quartic contributions from terms which are on the form
\bea
&&\<[T^{\aalpha},Y^{\bbeta};Y^{\ggamma}],[T^{\aalpha},Y^{\bbeta};Y^{\ggamma}]\> \cr
&=& 8 v^2 \lambda^4 \<[\phi^I,\phi^J][\phi_I,\phi_J] + 2[A_a,\phi^I][A^a,\phi_I] + [A_a,A_b][A^a,A^b]\>
\eea

\subsection{Summarizing}
We shall multiply the cubic term by a combinatorical factor of $12$ and the quartic term by a combinatorical factor of $3$. By collecting the terms, we have now found 
\bea
\<[XXX],[XXX]\>_{quadratic} &=& 24 \hbar^2 \lambda^2 \(\t f_{ab} \t f^{ab} + \partial_a \phi^I \partial^a \phi_I\)\cr
\<[XXX],[XXX]\>_{cubic} &=& 12.8 i v^2 \cepsilon \lambda^3 \(\frac{1}{2}\t f_{ab}[A^a,A^b] + \partial_a \phi^I [A^a,\phi_I]\)\cr
\<[XXX],[XXX]\>_{quartic} &=& 3.8v^2 \lambda^4 \([\phi^I,\phi^J][\phi_I,\phi_J] + 2[A_a,\phi^I][A^a,\phi_I] + [A_a,A_b][A^a,A^b]\)
\eea
Summing these terms we see that we obtain gauge covariant expressions. Reinstating the correct overall normalization, the total Lagrangian we have obtained up to zeroth order in $\frac{R}{v}$ is given by 
\bea
&&-\frac{1}{4g_{YM}^2} \Bigg(\frac{1}{4}\(\t F_{ab} \t F^{ab} + 2 F_{\mu a} F^{\mu a} + F_{\mu\nu} F^{\mu\nu}\) \cr
&&+ \frac{1}{2} \(D_a \phi^I D^a \phi_I + D_{\mu}\phi^I D^{\mu}\phi^I\) \cr
&&+ \frac{1}{4}[\phi_I,\phi_J][\phi^I,\phi^J] \cr
&& + \frac{1}{2v^2} \(G^{ab} A_a A_b + G_{IJ} \phi^I \phi^J  + 4 G \epsilon^{ab} \epsilon^{cd} v_a A_b \partial_c A_d\)\Bigg)
\eea
where
\bea
\t F_{ab} &=& F_{ab} + \cepsilon^{-1} \sqrt{G} \epsilon_{ab}\cr
F_{ab} &=& \partial_a A_b - \partial_b A_a - i[A_a,A_b]\cr
F_{\mu\nu} &=& \partial_{\mu}A_{\nu} - \partial_{\nu}A_{\mu} - i[A_{\mu},A_{\nu}]\cr
F_{\mu a} &=& \partial_{\mu} A_a - \partial_a A_{\mu} - i [A_{\mu},A_a]\cr
D_a \phi^I &=& \partial_a \phi^I - i[A_a,\phi^I]\cr
D_{\mu} \phi^I &=& \partial_{\mu} \phi^I - i [A_{\mu},\phi^I]
\eea
and 
\bea
g_{YM}^2 &=& 4\pi^2 \frac{v}{K}
\eea
Here we have thus rescaled the field so as to absorb the factor of $\frac{\cepsilon}{\lambda}$. 

Since we have generated gauge variant mass term for the gauge potential at order $\frac{1}{v}$ we see that our approximation where we cut out from a curved three-manifold, a flat two-torus, breaks down at this order. We should not trust this Lagrangian to this first order in $\frac{1}{v}$.

The inner product on the D4 brane Lagrangian has been suppressed. The inner product in ABJM theory is given by  
\bea
\frac{1}{N} \tr &=& \frac{1}{(2\pi)^2} \int d^2 \sigma
\eea
and we decompose 
\bea
N &=& N_{\A} N_{\B}
\eea
following \cite{Gustavsson:2010ep}. Then $N$ counts the number of M2 branes, $N_{\B}$ counts the number of $D4$ branes. After tracing over $\A$ indices, the residual the inner product on the D4 brane is over $\B$ indices,
\bea
\tr_{\B}
\eea
and is unit normalized since we start with the ABJM inner product which we decompose as 
\bea
\frac{1}{N_{\A}N_{\B}} \tr &=& \tr_{\B} \frac{1}{N_{\A}N_{\B}} \tr_{\A} 
\eea
Hence by mapping the star-product to matrix product in the D4 Lagrangian above, the inner product to be used is precisely $\tr_{\B}$.

\section{Single M5}
We saw that the D4 brane Lagrangian we obtained is not gauge covariant. To take proper care of the three-manifold which is rather invisible from D4 brane point of view, we will now instead consider the single M5 brane. Let us use real embedding coordinates $X^I$ as originally was used in BLG theory, for clarity. We then expand in fluctuations as
\bea
Y^I &=& Y^{\alpha} \partial_{\alpha} T^I
\eea
and ignore the scalar fields for the time being. Since $M_3$ is curved it is essential that we use covariant derivatives when computing the derivative
\bea
\partial_{\alpha} Y^I &=& D_{\alpha} Y^{\beta} \partial_{\beta} T^I + Y^{\beta} D_{\alpha} \partial_{\beta} T^I
\eea
Eventually we shall dualize
\bea
Y^{\alpha} &=& \frac{1}{2}\sqrt{g}\epsilon^{\alpha\beta\gamma}B_{\beta\gamma}
\eea
Our main goal now, is to in particular show that no gauge variant mass term $g_{\alpha\beta}Y^{\alpha}Y^{\beta}$ for the gauge potential arises when proper care is taken of the M5 brane geometry.

We expand the sextic potential to quadratic order. Let us here define the metric and the second fundamental form as
\bea
g_{\alpha\beta} &=& \partial_{\alpha}T^I \partial_{\beta} T^I\cr
\Omega_{\alpha\beta}^I &=& D_{\alpha} \partial_{\beta} T^I
\eea
By using the metric compatibility condition and the torsion free condition
\bea
D_{\gamma} g_{\alpha\beta} &=& 0\cr
\Omega_{\alpha\beta}^I &=& \Omega_{\beta\alpha}^I 
\eea
we obtain at quadratic order
\bea
-\frac{1}{12}\{X,X,X\}\{X,X,X\}|_{quadratic} &=& -\frac{1}{24}\(3(D_{\alpha}Y^{\alpha})^2 + M_{\alpha\beta} Y^{\alpha} Y^{\beta} -  Y^{\alpha} [D_{\alpha},D_{\beta}]Y^{\beta}\)
\eea
where
\bea
M_{\alpha\beta} &=& g^{\gamma\delta} \Omega_{\alpha\gamma}^I \Omega_{\beta\delta}^I
\eea
The first term is what we want. Upon dualizing it becomes
\bea
-\frac{3}{24}(D_{\alpha}Y^{\alpha})^2 &=& -\frac{1}{12}H_{\alpha\beta\gamma}H^{\alpha\beta\gamma}
\eea
where
\bea
H_{\alpha\beta\gamma} &=& D_{\alpha}B_{\beta\gamma}+D_{\gamma}B_{\alpha\beta}+D_{\beta}B_{\gamma\alpha}
\eea
The other two terms are unwanted and could give rise to gauge variant mass terms for the gauge potential. We will now demonstate the cancelation
\ben
Y^{\alpha}\(M_{\alpha\beta} - [D_{\alpha},D_{\beta}]\)Y^{\beta} &=& 0\label{cancel}
\een
We define the curvature tensor according to 
\bea
[D_{\alpha},D_{\beta}] V^{\gamma} &=& R^{\gamma}{}_{\tau\alpha\beta} V^{\tau}
\eea
We may use the Gauss-Codazzi relation\footnote{General expressions are found in \cite{Henningson:1999xi}.}
\bea
R_{\delta\gamma\alpha\beta} &=& \Omega_{\alpha\delta}^I \Omega_{\beta\gamma}^I - \Omega_{\alpha\gamma}^I \Omega_{\beta\delta}^I
\eea
from which follows that
\bea
R_{\alpha\beta} &=& M_{\alpha\beta} - \Omega_{\alpha\beta}^I \Omega^I
\eea
where we define
\bea
\Omega^I &=& g^{\alpha\beta} \Omega^I_{\alpha\beta}\cr
R_{\alpha\beta} &=& R^{\gamma}{}_{\alpha\beta\gamma}
\eea

We will now compute $\Omega^I$ explicitly for our specific three-manifold $M_3$. We switch to the complex basis where only $T^a$ are non-vanishing, and we compute
\bea
\Omega^a_{\alpha\beta} &=& \frac{1}{\sqrt{g}} \partial_{\alpha} \(\sqrt{g} g^{\alpha\beta} \partial_{\beta} T^a\)
\eea
for the embedding
\bea
T^a &=& x^a e^{i\psi}
\eea
and define $r = \sqrt{x^ax^a}$. Then
\bea
\Omega^b &=& \frac{1}{r}\partial_a r \partial_a T^b + \partial_a \partial_a T^b + \frac{1}{r^2} \partial_{\psi}^2 T^b\cr
&=& \(\frac{x^b}{r^2} - \frac{x^b}{r^2}\) e^{i\psi}\cr
&=& 0
\eea
for our specific three-manifold. By finally noting that
\bea
[D_{\alpha},D_{\beta}]Y^{\beta} &=& R_{\alpha\beta} Y^{\beta}
\eea
we see that the cancelation (\ref{cancel}) does occur. No gauge variant mass term for the gauge potential is generated.

\subsection{Direct derivation of the M5 brane coupling constant}
We have obtain the M5 brane coupling constant in a rather indirect way by obtaining the SYM coupling constant $g_{YM}^2 = 4\pi^2 v/K$, which is what we get by dimensional reduction of M5 brane on a circle of radius $v/K$. But it would also be nice if we could determine the M5 brane coupling constant directly by constructing the abelian M5 brane including the right normalization. We will address this problem here. 

After rescaling the scalar fields, our starting point is the Lagrangian
\bea
\frac{KN}{2\pi\hbar}\(-\frac{1}{2}\<D_{\mu}X^I,D^{\mu}X^I\> - \frac{1}{12} \<\{X^I,X^J,X^K\},\{X^I,X^J,X^K\}\>\)
\eea
where we focus only on the scalar fields. That will be sufficient for our purpose of determining the overall M5 brane coupling constant. In order to have a finite $\hbar$ we need to discretize the space. Again we do this by taking out a local two-torus at some large $v$ with radii $R$. Then we get as before
\bea
\hbar &=& \frac{2\pi R^2 v}{N}
\eea
Let us make the following ansatz for the fluctuation fields
\bea
Y^{\alpha} &=& \lambda \frac{1}{2} \sqrt{g} \epsilon^{\alpha\beta\gamma} B_{\beta\gamma}
\eea
and we express the inner product as
\bea
\<\> &=& \frac{1}{N} \tr\cr
&=& \int \frac{d^2 \sigma}{(2\pi)^2}\cr
&=& K\int \frac{d^2 \sigma d\psi}{(2\pi)^3} \cr
&=& \frac{K}{(2\pi)^3 R^2 v} \int d^3 \sigma \sqrt{g}
\eea
where $\sqrt{g} = R^2 v$ if the metric is $ds^2 = R^2 d\sigma^a d\sigma^a + v^2 d\psi^2$. We now see that by taking 
\ben
\lambda &=& \frac{4\pi^2 R^2 v}{KN}\label{lambda}
\een
the sextic potential gives the contribution
\bea
\frac{1}{4\pi} \(-\frac{1}{6} H_{\alpha\beta\gamma} H^{\alpha\beta\gamma}\)
\eea
to the full M5 brane Lagrangian. Other components of the three-form field strength, $H_{MNP}$ where $M=(\mu,\alpha)$, come from the Chern-Simons term and the kinetic term. But not even by taking all these contributions into account we get a fully Lorentz covariant expression. This is of course to be expected since $H_{MNP}$ is supposed to be selfdual and no Lorentz covariant action exists. But this problem is well-known and in the present case it has been analysed in \cite{Pasti:2009xc}. 

We obtain the correct normalization of the M5 brane Lagrangian for a two-form connection $H$ which is subject to the Dirac charge quantization 
\bea
\int_{3-cycle} \frac{1}{6} dx^M \wedge dx^N \wedge dx^P H_{MNP} &\in & 2\pi \mb{Z}
\eea
by choosing $\lambda$ as in (\ref{lambda}). Now it remains to explain this specific choice of $\lambda$. A natural three-cycle to consider in the present situation, is the two-torus over which we have a circle fiber. Then we have 
\bea
\int d^3\sigma \sqrt{g} \partial_{\alpha}Y^{\alpha} &=& \lambda \int d^3 \sigma \frac{1}{6} g\epsilon^{\alpha\beta\gamma}H_{\alpha\beta\gamma}
\eea
Now we have defined
\bea
Y^I &=& Y^{\alpha} \partial_{\alpha} T^I
\eea
it is natural to identify $Y^{\alpha}$ with a reparametrization of the torus,
\bea
\sigma^{\alpha} &\rightarrow & \sigma^{\alpha} + Y^{\alpha}
\eea
and then we must require
\bea
Y^{\alpha}(2\pi) & = & Y^{\alpha}(0) + 2\pi w^{\alpha}
\eea
where $w^{\alpha}$ are integer winding numbers. Then we get
\bea
\int d^3\sigma \sqrt{g} \partial_{\alpha}Y^{\alpha} &=& (2\pi)^3 R^2 v \(\frac{w^1}{K} + \frac{w^2}{K} + w^3\)
\eea
and from the Dirac quantization condition, we conclude that 
\bea
\lambda &=& \frac{4\pi^2 R^2 v}{K}
\eea
This misses out one factor of $N$ in the denominator. To get it, we would need the magnetic charge to be quantized as
\bea
\int H &=& 2\pi N \mb{Z}
\eea
Even though a similar magnetic charge has been observed for a compact D2 brane bound to $N$ D0 branes \cite{Taylor:2000za}, it appears to be different anyway. For $N$ D0 branes bound to a D2, the magnetic charge is given by $N$ and not $N\mb{Z}$.

\section{Discussion and outlook}
We have taken the infinite-$N$ solution of ABJM theory and discretized it by taking a small two-torus piece out of it. It might also be possible to consider the discretized solution which we may derive directly from ABJM theory, but for which only a series expansion in $1/v$ is presently known \cite{Terashima:2009fy}. Another issue we have largely omitted to discuss is the stability and possible supersymmetric extensions of this solution. One may also ask how to obtain SYM theories with other gauge groups than $U(N)$ gauge groups, from ABJM theories. This question seems to require a more general understanding of star-products and how they can be mapped into matrix multiplications.

\vskip 2truecm

\subsection*{Acknowledgements}
I would like to thank Futoshi Yagi for discussions. This work was supported by the National Research Foundation of Korea(NRF) grant funded by the Korea government(MEST) through the Center for Quantum Spacetime(CQUeST) of Sogang University with grant number 2005-0049409.


\newpage






\begin{thebibliography}{999}


\bibitem{Terashima:2009fy}
  S.~Terashima and F.~Yagi,
  ``M5-brane Solution in ABJM Theory and Three-algebra,''
  JHEP {\bf 0912} (2009) 059
  [arXiv:0909.3101 [hep-th]].

\bibitem{Terashima:2010ji}
  S.~Terashima and F.~Yagi,
  ``On Effective Action of Multiple M5-branes and ABJM Action,''
  arXiv:1012.3961 [hep-th].

\bibitem{Banks:1996nn}
  T.~Banks, N.~Seiberg and S.~H.~Shenker,
  ``Branes from matrices,''
  Nucl.\ Phys.\  B {\bf 490} (1997) 91
  [arXiv:hep-th/9612157].


\bibitem{Gustavsson:2010ep}
  A.~Gustavsson,
  ``Five-dimensional super Yang-Mills theory from ABJM theory,''
  arXiv:1012.5917 [hep-th].


\bibitem{Gustavsson:2010nc}
  A.~Gustavsson,
  ``An associative star-three-product and applications to M two/M five-brane
  theory,''
  JHEP {\bf 1011}, 043 (2010)
  [arXiv:1008.0902 [hep-th]].

\bibitem{Bagger:2008se}
  J.~Bagger and N.~Lambert,
  ``Three-Algebras and N=6 Chern-Simons Gauge Theories,''
  Phys.\ Rev.\  D {\bf 79}, 025002 (2009)
  [arXiv:0807.0163 [hep-th]].

\bibitem{Nastase:2009ny}
  H.~Nastase, C.~Papageorgakis and S.~Ramgoolam,
  ``The fuzzy $S^2$ structure of M2-M5 systems in ABJM membrane theories,''
  JHEP {\bf 0905} (2009) 123
  [arXiv:0903.3966 [hep-th]].


\bibitem{Henningson:1999xi}
  M.~Henningson and K.~Skenderis,
  ``Weyl anomaly for Wilson surfaces,''
  JHEP {\bf 9906}, 012 (1999)
  [arXiv:hep-th/9905163].

\bibitem{Pasti:2009xc}
  P.~Pasti, I.~Samsonov, D.~Sorokin and M.~Tonin,
  ``BLG-motivated Lagrangian formulation for the chiral two-form gauge field in
  D=6 and M5-branes,''
  Phys.\ Rev.\  D {\bf 80}, 086008 (2009)
  [arXiv:0907.4596 [hep-th]].

\bibitem{Taylor:2000za}
  W.~Taylor,
  ``D2-branes in B fields,''
  JHEP {\bf 0007}, 039 (2000)
  [arXiv:hep-th/0004141].

\bibitem{Taylor:2001vb}
  W.~Taylor,
  ``M(atrix) theory: Matrix quantum mechanics as a fundamental theory,''
  Rev.\ Mod.\ Phys.\  {\bf 73}, 419 (2001)
  [arXiv:hep-th/0101126].


 


\end{thebibliography}
\end{document}